\documentclass[a4paper]{article}

\usepackage{icrc2013}
\usepackage[english]{babel}
\title{A new 3D transport and radiation code for galactic cosmic rays}

\shorttitle{A new 3D transport and radiation code for galactic cosmic rays}

\authors{
M. Werner$^{1}$,
R. Kissmann$^{1}$,
A. W. Strong$^{2}$,
O. Reimer$^{1}$
}

\afiliations{
$^1$ Institut f\"ur Astro- und Teilchenphysik, Leopold-Franzens-Universit\"at Innsbruck, A-6020 Innsbruck, Austria \\
$^2$ Max Planck Institut f\"ur extraterrestrische Physik, Postfach 1312, D-85741 Garching, Germany \\
}

\email{Michael.Werner@uibk.ac.at}

\abstract{We show the necessity for a new approach towards
  comprehensive and consistent simulations of the propagation of
  galactic cosmic rays. Our developments are optimised for addressing
  the spatially 3-dimensional inhomogeneous diffusion problem and
  utilise contemporary numerical methods. We aim to address the
  transport problem in a full 3-dimensional environment. For that, we
  test the transition from 2D to 3D simulation results within an
  existing propagation code.  We present sub-kpc scale simulations
  that allow the investigation of small-scale structures regarding
  different model conditions such as variety regarding
  non-axisymmetric cosmic ray source distributions. These results are
  discussed critically and motivate our development of a new transport
  code for galactic cosmic rays. The capabilities of this code are
  outlined.}


\begin{document}
\maketitle

\section{Introduction}
The challenge in modelling the propagation of cosmic rays (CRs) in our
Galaxy is at least twofold: Implementing a realistic and accordingly
complex physics model of CR propagation in our Galaxy and devising a
numerical scheme that can solve the underlying transport equations
accurately but efficiently. These two conflicting aspects are often
consolidated by making several more or less justified simplifications
in the underlying CR propagation physics and/or the input parameter
space (i.e. galactic magnetic field, gas distribution, radiation
fields, isotropic diffusion). Consequently, current available
transport codes feature CR propagation models of varying complexity
\cite{bib:review}, from single species fluid models to codes that
account for a nuclear reaction network of multiple CR species. DRAGON
\cite{bib:dragon1} and GALPROP \cite{bib:galprop} are considered the
most capable of the currently publicly available codes. Both compute
the CR distribution in our Galaxy by solving the CR transport equation
for each CR species using a mostly second-order Crank-Nicolson
discretisation. The most prominent CR interactions with the radiation
fields and matter distributions in our Galaxy are taken into account,
furthermore both allow the computation of secondary particles,
e.g. $\gamma$-rays. Gamma-rays are of particular interest as they
offer a testing ground for the validity of CR particle
transport modelling results via comparison to measurements of the
galactic diffuse $\gamma$-ray emission.  

 \begin{figure}[ht!]
  \centering
  \includegraphics[width=0.45\textwidth]{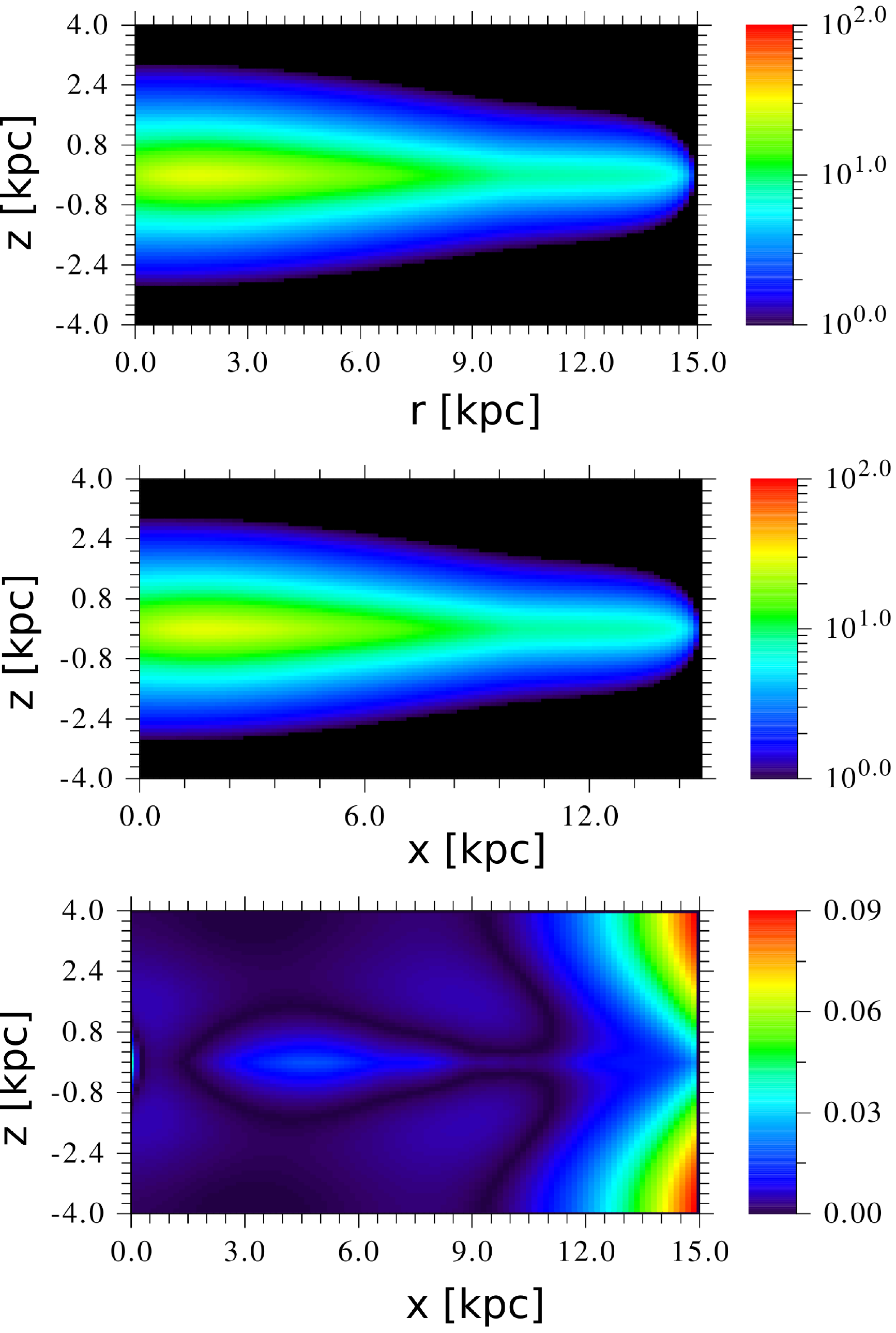}
  \caption{Above: $E_{kin}$ = 100 MeV proton density distribution as
    obtained by a 2D propagation scenario. Middle: x,z-slice of
    $E_{kin}$ = 100 MeV proton density distribution obtained by a 3D propagation scenario. Below: Residual ($ | \frac{2D - 3D}{3D} |$) of the two proton distributions shown above. Colour scale of CR density in arbitrary units.}
  \label{fig01}
 \end{figure}

Despite their capabilities current propagation scenarios are
still relying on simplifications that have been rendered obsolete by
tighter constraint on the input parameters as provided by new
experimental results, i. e. matter distributions, magnetic field
models, radiation fields, CR source distributions, and perhaps more
importantly by the increase in computational power that allows a more
realistic 3-dimensional non-isotropic treatment of CR
propagation. Only recently steps have been made to overcome the
2D-paradigm that clearly demonstrate that new insights on current
topics in astroparticle physics can be gained \cite{bib:dragon2}.

We show if and how 3-dimensional sub-kpc scale propagations scenarios
can be treated using GALPROP and test the validity by comparing these
results to 2-dimensional simulations. We demonstrate the capabilities of 3-dimensional simulations by
introducing a non-axisymmetric CR source distribution. In this context
we discuss current limitations of GALPROP thereby motivating the
necessity of new code developments.


\section{Beyond the 2D-paradigm}
\subsection{2D to 3D comparison}
\begin{figure}[ht]
  \centering
  \includegraphics[width=0.48\textwidth]{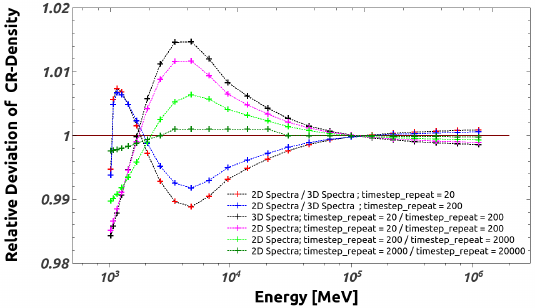}
  \caption{Comparison of CR proton spectra obtained from 2D and 3D scenarios: for \texttt{timestep\_repeat} = 20 (\textit{red crosses}) and 200 (\textit{blue crosses}). Deviations of 2D solutions: for \texttt{timestep\_repeat} = 20 to = 200 (\textit{magenta crosses}), 200 to 2000 (\textit{light green crosses}) and 2000 to 20000 (\textit{dark green crosses}) as well as for 3D solution using \texttt{timestep\_repeat} = 20 compared to = 200 (\textit{black crosses}).}
  \label{fig02}
 \end{figure}

We test consistency of the solution for a 3D-propagation scenario
obtained by GALPROP with that obtained in the 2D case by formulating a
2D ($z$,$r$-coordinates) scenario using an axisymmetric source
distribution (\texttt{source\_model=1}) and an equivalent 3D
propagation scenario that should in principle result in the same CR
distribution. We use a spatial resolution of 0.1 kpc for the
z-direction and 0.15 kpc for the r,x,y-directions. In these scenarios
our Galaxy is confined within a box ranging from x,y = $\pm 15$ kpc (r
= 15kpc in the 2D scenario) and a height ranging from z = $\pm 4$
kpc. Here we discuss protons and electrons only, although tests with
nuclei up to Z = 8 have been performed. The energy grid uses 23
logarithmically equidistant energy points ranging from 100 MeV to
about 1 TeV. The size of the time steps used by the solver ranges from
$10^{8}$ yrs to $10^{2}$ yrs.\footnote{The corresponding
  \texttt{GALDEF} files are available upon request}. We find that the
2D and 3D proton distributions match and that any deviations are on
the percent level. This is exemplary shown for $E_{kin} = $ 100 MeV
protons in Figure \ref{fig01}. Up to a certain degree of accuracy,
GALPROP provides a consistent solution. We use the CR proton spectrum
at Earth to quantify the remaining discrepancies between the 2D and 3D
solutions and investigate their dependence on parameters that govern
the numerical solver. This is shown exemplary for the parameter
\texttt{timestep\_repeat} (the number of iterations in each time step)
in Figure \ref{fig02}.
In case of CR proton spectra the 2D and 3D scenarios seem to converge
toward a common solution. We note however that in case of the 3D scenario and
for the number of iterations exceeding 200, the required computing time
becomes excessive. Figure \ref{fig02} also shows that the relative
deviations of the solutions obtained in the 2D scenario decrease with
increasing number of iterations, a further indication of
convergence. The same holds true for the 3D scenario. These test are
necessary because GALPROP does not properly control convergence
towards a numerical solution nor offers feedback on the error of the
particular solution.
 \begin{figure}[h!]
  \centering
  \includegraphics[width=0.48\textwidth]{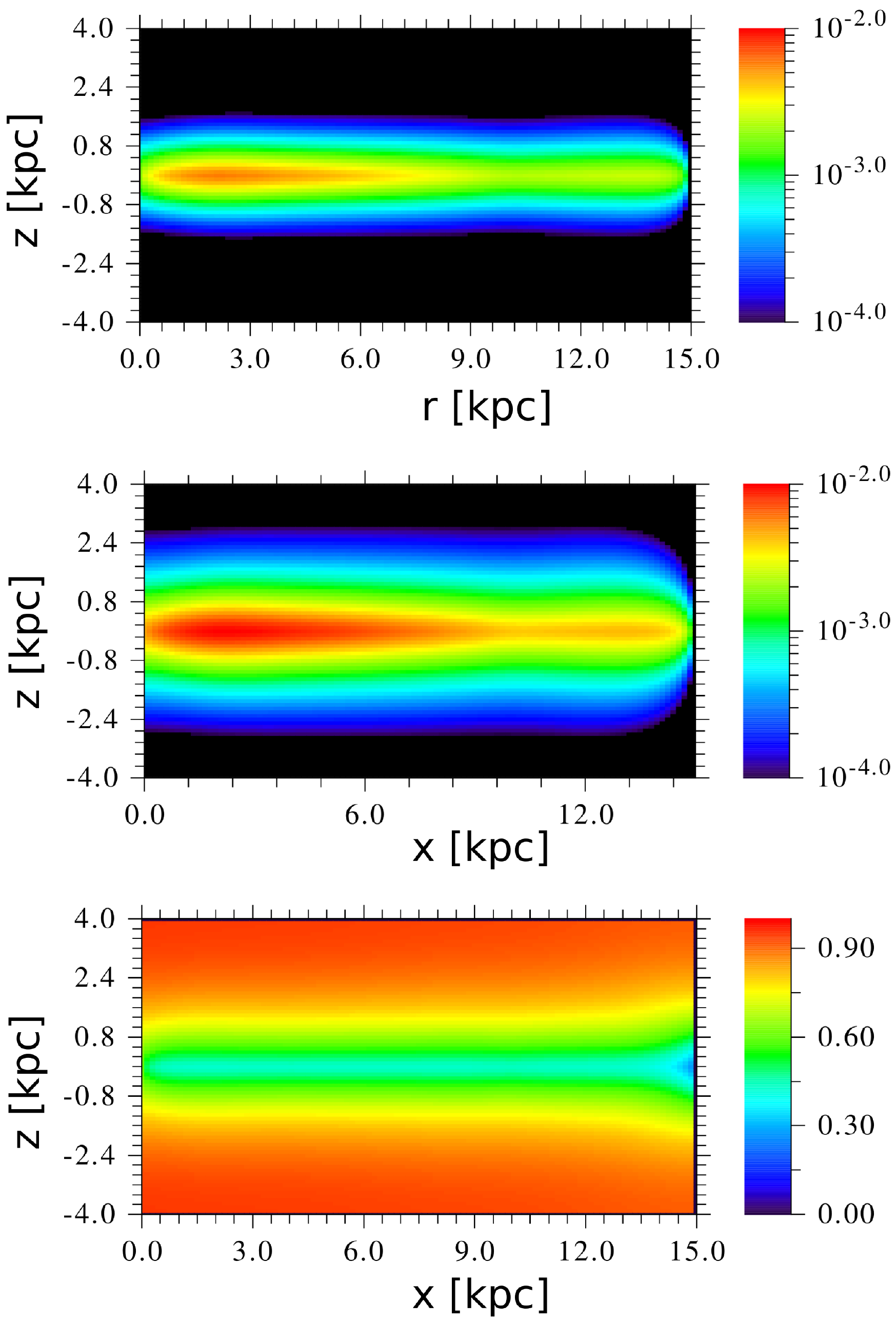}
  \caption{Above: $E_{kin}$ = 1.2 TeV electron density distribution as
    obtained by a 2D propagation scenario. Middle: x,z-slice of the
    $E_{kin}$ = 1.2 TeV electron density distribution obtained by a 3D
    propagation scenario. Below: Residual ($ | \frac{2D - 3D}{3D} |$) of the two electron distributions shown above. Colour scale of CR density in arbitrary units.}
  \label{fig03}
 \end{figure}
 
  \begin{figure}[t]
  \centering
  \includegraphics[width=0.48\textwidth]{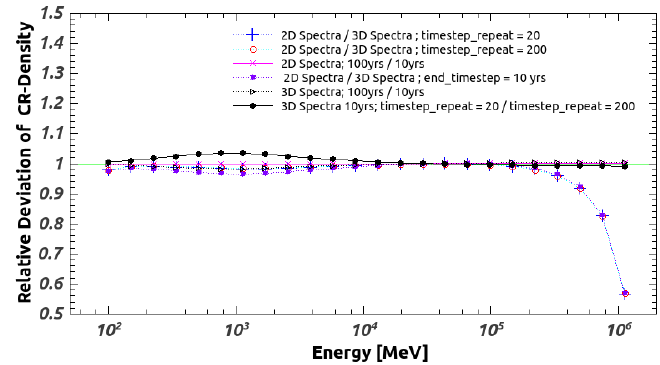}
  \caption{Comparison of CR electron spectra at Earth for: Spectra derived from 2D scenario divided by spectra obtained from 3D simulations for \texttt{timestep\_repeat} = 20 (\textit{blue crosses}), = 200 (\textit{red circle} and \texttt{end\_timestep} = 10 yrs (\textit{purple star})). Comparison of 2D spectra obtained with \texttt{end\_timestep} = 100 yrs to = 10 yrs (\textit{magenta crosses}) and analogously for 3D (\textit{black arrows}) and using \texttt{end\_timestep} = 10 yrs (\textit{black circle})}
  \label{fig04}
 \end{figure}
 
  \begin{figure}[t]
  \centering
  \includegraphics[width=0.48\textwidth]{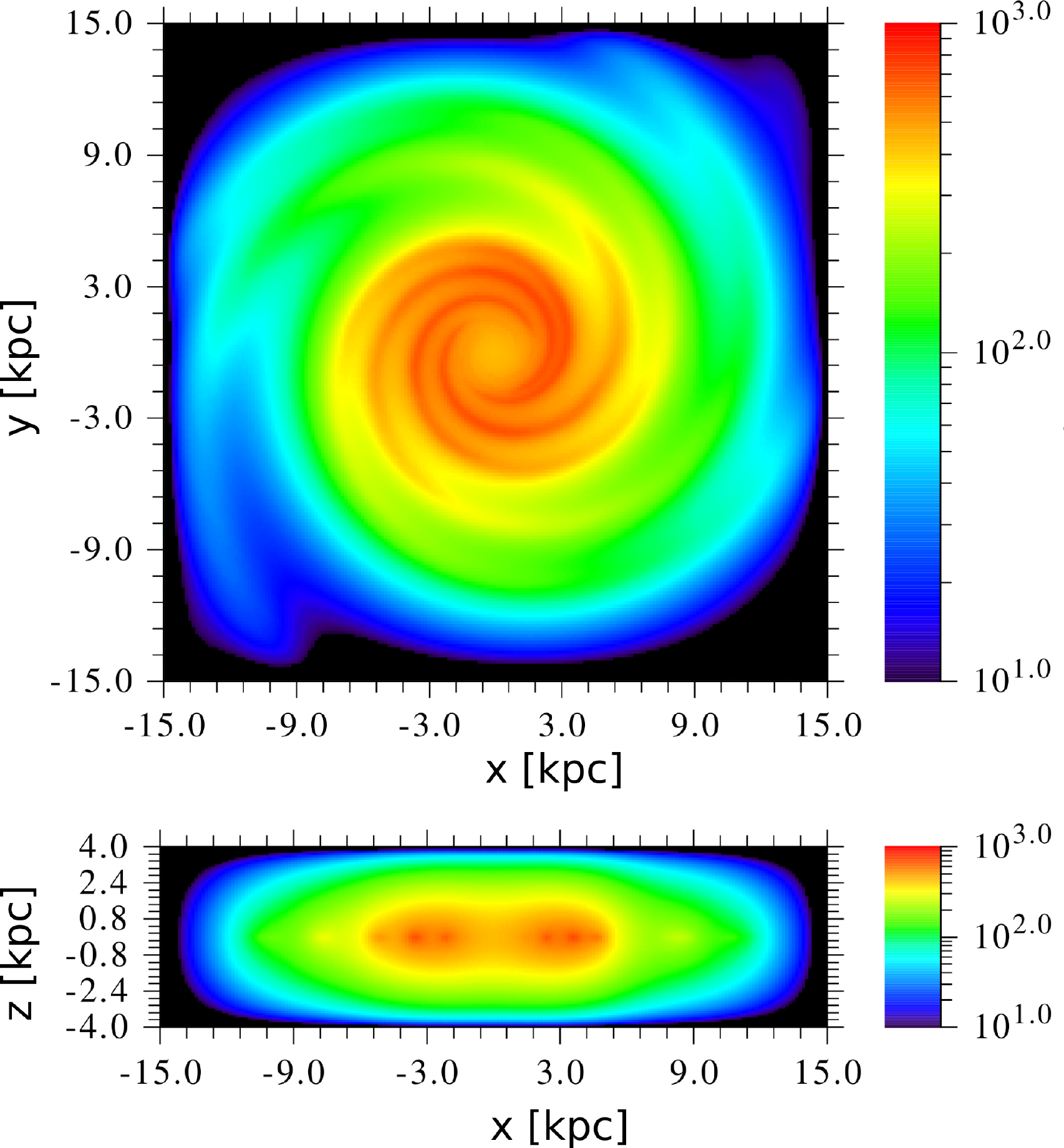}
  \caption{Above: x,y-slice at z = 0 of the proton density distribution at an energy of $E_{kin} = 444$ MeV obtained using a CR source distribution following a spiral arm model. Below: y,z-slice at x = 0 of the same distribution as above. Colour scale of CR density in arbitrary units.}
  \label{fig05}
 \end{figure}
 
   \begin{figure}[ht!]
  \centering
  \includegraphics[width=0.48\textwidth]{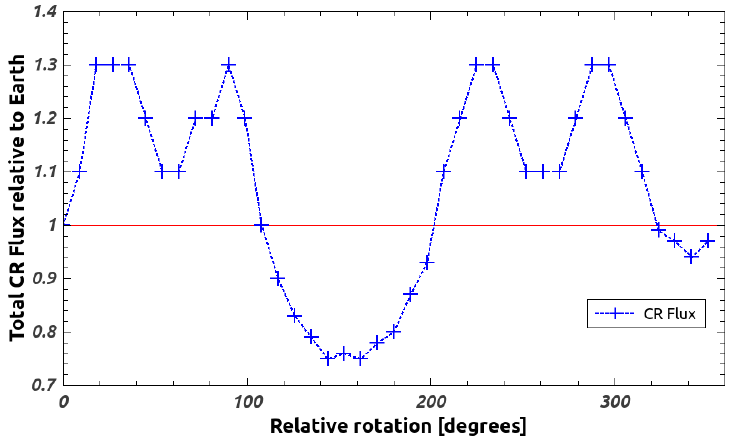}
  \caption{Total CR flux at Earth for different positions of Earth relative to the spiral arm pattern. Flux is given relative to the flux at Earth's initial position (8.5,0,0). The x-axis denotes the counter clock wise rotation angle}
  \label{fig06}
 \end{figure}
We performed a similar analysis for electrons. As an example we show
the spatial electron distribution at an energy of $E_{kin} = $ 1.2 TeV
in Figure \ref{fig03} where deviations are considerably larger. In
Figure \ref{fig04} we show that the solutions of the 2D and 3D
scenarios for energies above 200 GeV do not match as well as for CR
protons. We use the electron spectra at Earth to quantify the
deviations and find that the deviations increase with increasing
energy. We do not find any dependence on the number of iterations in
each time step nor the size of the smallest time step. If this
discrepancy hints at an error in the numerics, e.g., that the high
energy boundary conditions are handled incorrectly, or that the time
scales associated with energy loss are simply too small, is still
under investigation.

\subsection{3D modelling}
3D-simulations allow us to model propagation scenarios that do not
feature $\phi$-symmetry. To demonstrate this we implemented a source
distribution following a logarithmic spiral arm pattern derived from
COBE observations of FIR cooling lines \cite{bib:spiral}. 
Figure \ref{fig05} shows the resulting distribution of CR protons at an energy of $E_{kin} = 444$
MeV. This allows us to study a range of new aspects inaccessible by 2D
scenarios, for instance the significant change of the CR flux at Earth with Earth's position relative to the spiral arms. This is shown in Figure \ref{fig06}.

\begin{figure}[t]
  \centering
  \includegraphics[width=0.48\textwidth]{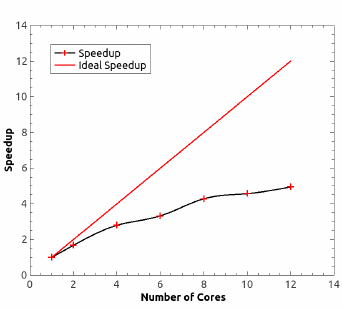}
  \caption{Speedup of GALPROP on a shared memory machine using the OpenMP API.}
  \label{fig07}
 \end{figure}

For all its capabilities and its long history of development heritage
GALPROP has several shortcomings which demand improvement of the
existing or the development of a new CR transport code. Higher
resolutions require the code to perform well on parallel high
performance computer architectures. GALPROP is capable of running on
multiple cores on shared memory systems using the OpenMP API but, as
we show in Figure \ref{fig07}, the speed-up gained by using additional
computing cores is far from optimal. We note that in most cases the
limiting factor is the available memory. Currently, GALPROP cannot
take advantage of distributed memory machines, thus, high spatial
resolution simulations are impossible on such computing
infrastructures. GALPROP also does not allow a quantification of the
numerical error and convergence towards a solution has to be
determined by trial and error \cite{bib:galprop}. GALPROP relies on
the assumption that CR diffusion is isotropic and uses a scalar for
the diffusion coefficient rather than a tensor. A recent study has
shown that abandoning isotropic diffusion has a substantial effect on
the CR spectrum at Earth \cite{bib:aniso}. A further limitation is
that GALPROP calculates the nuclear reaction network after it obtains
a propagation solution and is therefore unable to include time-dependent effects of the nuclear reaction network correctly.

\section{A new CR transport code}
In the previous section we identified the lack of control over
convergence in the GALPROP code as a fundamental problem for the
numerical solver. While we showed that it is actually possible to find
a meaningful quantification of the convergence of the code, the less
experienced user of such a code, is hardly interested in repeatedly
conducting convergence studies between any major change in the
physical setup.

This problem is actually connected to the way the transport equation
is solved within GALPROP: the transport equation is integrated in time
until the user-specified total propagation time is reached. In this
case the user has to make sure that the overall time integrated by the
solver is sufficiently long for a steady state solution to be
achieved. An obvious alternative to this scheme would be to compute a
steady-state solution directly without any time-integration. This,
however, forbids a time-splitting approach as utilised within GALPROP:
a first order operator splitting method is applied in a way that all
dimensions are alternately solved for, while keeping the remaining
coordinate dimensions fixed.

Therefore, we are deploying an alternative solver which no longer uses
the dimensional operator splitting. This of course has the drawback
that for a Crank-Nicolson discretisation of the diffusion problem the
resulting matrix-equation no longer has tri-diagonal form. While the
tri-diagonal form has the advantage that a solution can be computed
directly and efficiently, modern numerical methods can handle sparse
matrix problems very efficiently, too. We are currently testing the
implementation of such a method. In particular the solution of the
steady state problem is very efficient as it does not invoke any
time-integration steps. Here we are using two different approaches in
parallel. On the one hand we implemented a method, which solves the
steady state problem only. This has the advantage that there is no
longer a decrease in accuracy due to the order of the time
discretisation. On the other hand we also implemented a time-dependent
solver, where we can follow the time-dependent evolution of the Cosmic
Ray distribution within the Galaxy. This is, e.g., used in cases where
temporally variable sources are needed to be considered. For the
latter we use the result from the steady state solver as an initial
condition that is to be modified by temporal effects. The new solver
uses operator splitting for the time-dependent problem.  We use
adapted solvers for each \emph{physically} distinct term in the
transport equation instead of simple dimensional operator splitting as
used within GALPROP. That is we apply different solvers for spatial
convection, spatial diffusion and diffusion and energy losses in
momentum space.  Within the solution of the transport problem we allow
for a full spatial variation of the diagonal components of the
diffusion tensor. Corresponding analytical tests show satisfactory
results.

A specific description of these new solvers will be given in an
upcoming publication where we will show the convergence properties and
the numerical error by comparison to analytical tests. In particular,
the final result will no longer rely on convergence criteria to be
selected by the user. This allows the user to concentrate on the
physical problem at hand and facilitate consistent comparisons within
the community. Our new CR transport code is capable of utilizing
modern parallel computing architecture by using the Message Passing
Interface (MPI) standard.

\section{Summary and Discussion}
New observational constraints on the input parameters of CR transport
models as well as the increase in available computing power
necessitate a transitions from 2D modelling towards 3D simulations of
CR propagation in our Galaxy. We study this transition from 2D to 3D
simulations using an existing code (GALPROP), and find significant
deviations for electrons with energies higher than 200 GeV. For
protons we find a general agreement except for discrepancies on the
percent level. Our comparison shows the need for practical and
universal convergence criteria. Using sub-kpc scale 3D simulations we
investigate a non-axisymmetric CR source distributions following a
spiral arm pattern that allows us to address scientific questions
inaccessible to 2D simulations. We demonstrate the need for an
efficient CR transport code that makes use of parallel computing
architectures.

Motivated by our findings we present a brief overview of our new CR
transport code. This code uses contemporary numerical methods in a way
that the user no longer unknowingly faces convergence issues. The new
code will rely on a number of solvers, each optimized for different
physical sub-processes in the CR propagation equation, and is capable
of treating anisotropic diffusion. Our CR transport code utilises
modern parallel computing architectures.

\vspace*{0.5cm}
\footnotesize{{\bf Acknowledgement:}{The work presented in this paper was supported by the Austrian Science Fund FWF}. Part of this work was done while M. Werner was guest at the Max Planck Institute for Extraterrestrial Physics.}

\end{document}